# Tests of QCD in $W$ and $Z$ Production at Tevatron*

The DØ Collaboration

## Abstract

We present measurements of the production cross sections times leptonic branching fractions and the transverse momentum distributions of $W$ and $Z$ bosons in $p\bar{p}$ collisions at $\sqrt{s} = 1.8$ TeV using data collected with the DØ detector at the Fermilab Tevatron $p\bar{p}$ collider. A preliminary measurement of the $W$ charge asymmetry is also presented.

S. Abachi,[12] B. Abbott,[34] M. Abolins,[23] B.S. Acharya,[41] I. Adam,[10] D.L. Adams,[35] M. Adams,[15] S. Ahn,[12] H. Aihara,[20] J. Alitti,[37] G. Álvarez,[16] G.A. Alves,[8] E. Amidi,[27] N. Amos,[22] E.W. Anderson,[17] S.H. Aronson,[3] R. Astur,[39] R.E. Avery,[29] A. Baden,[21] V. Balamurali,[30] J. Balderston,[14] B. Baldin,[12] J. Bantly,[4] J.F. Bartlett,[12] K. Bazizi,[7] J. Bendich,[20] S.B. Beri,[32] I. Bertram,[35] V.A. Bezzubov,[33] P.C. Bhat,[12] V. Bhatnagar,[32] M. Bhattacharjee,[11] A. Bischoff,[7] N. Biswas,[30] G. Blazey,[12] S. Blessing,[13] P. Bloom,[5] A. Boehnlein,[12] N.I. Bojko,[33] F. Borcherding,[12] J. Borders,[36] C. Boswell,[7] A. Brandt,[12] R. Brock,[23] A. Bross,[12] D. Buchholz,[29] V.S. Burtovoi,[33] J.M. Butler,[12] D. Casey,[36] H. Castilla-Valdez,[9] D. Chakraborty,[39] S.-M. Chang,[27] S.V. Chekulaev,[33] L.-P. Chen,[20] W. Chen,[39] L. Chevalier,[37] S. Chopra,[32]

B.C. Choudhary,[7] J.H. Christenson,[12] M. Chung,[15] D. Claes,[39] A.R. Clark,[20] W.G. Cobau,[21]
J. Cochran,[7] W.E. Cooper,[12] C. Cretsinger,[36] D. Cullen-Vidal,[4] M.A.C. Cummings,[14] D. Cutts,[4]
O.I. Dahl,[20] K. De,[42] M. Demarteau,[12] R. Demina,[27] K. Denisenko,[12] N. Denisenko,[12]
D. Denisov,[12] S.P. Denisov,[33] W. Dharmaratna,[13] H.T. Diehl,[12] M. Diesburg,[12] G. Di Loreto,[23]
R. Dixon,[12] P. Draper,[42] J. Drinkard,[6] Y. Ducros,[37] S.R. Dugad,[41] S. Durston-Johnson,[36]
D. Edmunds,[23] J. Ellison,[7] V.D. Elvira,[12,‡] R. Engelmann,[39] S. Eno,[21] G. Eppley,[35] P. Ermolov,[24]
O.V. Eroshin,[33] V.N. Evdokimov,[33] S. Fahey,[23] T. Fahland,[4] M. Fatyga,[3] M.K. Fatyga,[36]
J. Featherly,[3] S. Feher,[39] D. Fein,[2] T. Ferbel,[36] G. Finocchiaro,[39] H.E. Fisk,[12] Yu. Fisyak,[24]
E. Flattum,[23] G.E. Forden,[2] M. Fortner,[28] K.C. Frame,[23] P. Franzini,[10] S. Fuess,[12]
A.N. Galjaev,[33] E. Gallas,[42] C.S. Gao,[12,*] S. Gao,[12,*] T.L. Geld,[23] R.J. Genik II,[23] K. Genser,[12]
C.E. Gerber,[12,§] B. Gibbard,[3] V. Glebov,[36] S. Glenn,[5] B. Gobbi,[29] M. Goforth,[13]
A. Goldschmidt,[20] B. Gómez,[1] P.I. Goncharov,[33] H. Gordon,[3] L.T. Goss,[43] N. Graf,[3]
P.D. Grannis,[39] D.R. Green,[12] J. Green,[28] H. Greenlee,[12] G. Griffin,[6] N. Grossman,[12]
P. Grudberg,[20] S. Grünendahl,[36] W. Gu,[12,*] G. Guglielmo,[31] J.A. Guida,[39] J.M. Guida,[3]
W. Guryn,[3] S.N. Gurzhiev,[33] P. Gutierrez,[31] Y.E. Gutnikov,[33] N.J. Hadley,[21] H. Haggerty,[12]
S. Hagopian,[13] V. Hagopian,[13] K.S. Hahn,[36] R.E. Hall,[6] S. Hansen,[12] R. Hatcher,[23]
J.M. Hauptman,[17] D. Hedin,[28] A.P. Heinson,[7] U. Heintz,[12] R. Hernández-Montoya,[9]
T. Heuring,[13] R. Hirosky,[13] J.D. Hobbs,[12] B. Hoeneisen,[1,¶] J.S. Hoftun,[4] F. Hsieh,[22] Ting Hu,[39]
Tong Hu,[16] T. Huehn,[7] S. Igarashi,[12] A.S. Ito,[12] E. James,[2] J. Jaques,[30] S.A. Jerger,[23]
J.Z.-Y. Jiang,[39] T. Joffe-Minor,[29] H. Johari,[27] K. Johns,[2] M. Johnson,[12] H. Johnstad,[40]
A. Jonckheere,[12] M. Jones,[14] H. Jöstlein,[12] S.Y. Jun,[29] C.K. Jung,[39] S. Kahn,[3] G. Kalbfleisch,[31]
J.S. Kang,[18] R. Kehoe,[30] M.L. Kelly,[30] A. Kernan,[7] L. Kerth,[20] C.L. Kim,[18] S.K. Kim,[38]
A. Klatchko,[13] B. Klima,[12] B.I. Klochkov,[33] C. Klopfenstein,[39] V.I. Klyukhin,[33]
V.I. Kochetkov,[33] J.M. Kohli,[32] D. Koltick,[34] A.V. Kostritskiy,[33] J. Kotcher,[3] J. Kourlas,[26]
A.V. Kozelov,[33] E.A. Kozlovski,[33] M.R. Krishnaswamy,[41] S. Krzywdzinski,[12] S. Kunori,[21]
S. Lami,[39] G. Landsberg,[12] R.E. Lanou,[4] J-F. Lebrat,[37] A. Leflat,[24] H. Li,[39] J. Li,[42] Y.K. Li,[29]
Q.Z. Li-Demarteau,[12] J.G.R. Lima,[8] D. Lincoln,[22] S.L. Linn,[13] J. Linnemann,[23] R. Lipton,[12]
Y.C. Liu,[29] F. Lobkowicz,[36] S.C. Loken,[20] S. Lökös,[39] L. Lueking,[12] A.L. Lyon,[21] A.K.A. Maciel,[8]

R.J. Madaras,[20] R. Madden,[13] I.V. Mandrichenko,[33] Ph. Mangeot,[37] S. Mani,[5] B. Mansoulié,[37] H.S. Mao,[12,*] S. Margulies,[15] R. Markeloff,[28] L. Markosky,[2] T. Marshall,[16] M.I. Martin,[12] M. Marx,[39] B. May,[29] A.A. Mayorov,[33] R. McCarthy,[39] T. McKibben,[15] J. McKinley,[23] T. McMahon,[31] H.L. Melanson,[12] J.R.T. de Mello Neto,[8] K.W. Merritt,[12] H. Miettinen,[35] A. Milder,[2] A. Mincer,[26] J.M. de Miranda,[8] C.S. Mishra,[12] M. Mohammadi-Baarmand,[39] N. Mokhov,[12] N.K. Mondal,[41] H.E. Montgomery,[12] P. Mooney,[1] M. Mudan,[26] C. Murphy,[16] C.T. Murphy,[12] F. Nang,[4] M. Narain,[12] V.S. Narasimham,[41] A. Narayanan,[2] H.A. Neal,[22] J.P. Negret,[1] E. Neis,[22] P. Nemethy,[26] D. Nešić,[4] D. Norman,[43] L. Oesch,[22] V. Oguri,[8] E. Oltman,[20] N. Oshima,[12] D. Owen,[23] P. Padley,[35] M. Pang,[17] A. Para,[12] C.H. Park,[12] Y.M. Park,[19] R. Partridge,[4] N. Parua,[41] M. Paterno,[36] J. Perkins,[42] A. Peryshkin,[12] M. Peters,[14] H. Piekarz,[13] Y. Pischalnikov,[34] A. Pluquet,[37] V.M. Podstavkov,[33] B.G. Pope,[23] H.B. Prosper,[13] S. Protopopescu,[3] D. Pušeljić,[20] J. Qian,[22] P.Z. Quintas,[12] R. Raja,[12] S. Rajagopalan,[39] O. Ramirez,[15] M.V.S. Rao,[41] P.A. Rapidis,[12] L. Rasmussen,[39] A.L. Read,[12] S. Reucroft,[27] M. Rijssenbeek,[39] T. Rockwell,[23] N.A. Roe,[20] P. Rubinov,[39] R. Ruchti,[30] S. Rusin,[24] J. Rutherfoord,[2] A. Santoro,[8] L. Sawyer,[42] R.D. Schamberger,[39] H. Schellman,[29] J. Sculli,[26] E. Shabalina,[24] C. Shaffer,[13] H.C. Shankar,[41] R.K. Shivpuri,[11] M. Shupe,[2] J.B. Singh,[32] V. Sirotenko,[28] W. Smart,[12] A. Smith,[2] R.P. Smith,[12] R. Snihur,[29] G.R. Snow,[25] S. Snyder,[39] J. Solomon,[15] P.M. Sood,[32] M. Sosebee,[42] M. Souza,[8] A.L. Spadafora,[20] R.W. Stephens,[42] M.L. Stevenson,[20] D. Stewart,[22] D.A. Stoianova,[33] D. Stoker,[6] K. Streets,[26] M. Strovink,[20] A. Taketani,[12] P. Tamburello,[21] J. Tarazi,[6] M. Tartaglia,[12] T.L. Taylor,[29] J. Teiger,[37] J. Thompson,[21] T.G. Trippe,[20] P.M. Tuts,[10] N. Varelas,[23] E.W. Varnes,[20] P.R.G. Virador,[20] D. Vititoe,[2] A.A. Volkov,[33] A.P. Vorobiev,[33] H.D. Wahl,[13] G. Wang,[13] J. Wang,[12,*] L.Z. Wang,[12,*] J. Warchol,[30] M. Wayne,[30] H. Weerts,[23] F. Wen,[13] W.A. Wenzel,[20] A. White,[42] J.T. White,[43] J.A. Wightman,[17] J. Wilcox,[27] S. Willis,[28] S.J. Wimpenny,[7] J.V.D. Wirjawan,[43] J. Womersley,[12] E. Won,[36] D.R. Wood,[12] H. Xu,[4] R. Yamada,[12] P. Yamin,[3] C. Yanagisawa,[39] J. Yang,[26] T. Yasuda,[27] C. Yoshikawa,[14] S. Youssef,[13] J. Yu,[36] Y. Yu,[38] Y. Zhang,[12,*] Y.H. Zhou,[12,*] Q. Zhu,[26] Y.S. Zhu,[12,*] Z.H. Zhu,[36] D. Zieminska,[16] A. Zieminski,[16] and A. Zylberstejn[37]

[1]Universidad de los Andes, Bogotá, Colombia

[2]University of Arizona, Tucson, Arizona 85721

[3]Brookhaven National Laboratory, Upton, New York 11973

[4]Brown University, Providence, Rhode Island 02912

[5]University of California, Davis, California 95616

[6]University of California, Irvine, California 92717

[7]University of California, Riverside, California 92521

[8]LAFEX, Centro Brasileiro de Pesquisas Físicas, Rio de Janeiro, Brazil

[9]CINVESTAV, Mexico City, Mexico

[10]Columbia University, New York, New York 10027

[11]Delhi University, Delhi, India 110007

[12]Fermi National Accelerator Laboratory, Batavia, Illinois 60510

[13]Florida State University, Tallahassee, Florida 32306

[14]University of Hawaii, Honolulu, Hawaii 96822

[15]University of Illinois at Chicago, Chicago, Illinois 60607

[16]Indiana University, Bloomington, Indiana 47405

[17]Iowa State University, Ames, Iowa 50011

[18]Korea University, Seoul, Korea

[19]Kyungsung University, Pusan, Korea

[20]Lawrence Berkeley Laboratory and University of California, Berkeley, California 94720

[21]University of Maryland, College Park, Maryland 20742

[22]University of Michigan, Ann Arbor, Michigan 48109

[23]Michigan State University, East Lansing, Michigan 48824

[24]Moscow State University, Moscow, Russia

[25]University of Nebraska, Lincoln, Nebraska 68588

[26]New York University, New York, New York 10003

[27]Northeastern University, Boston, Massachusetts 02115

[28]Northern Illinois University, DeKalb, Illinois 60115

[29]Northwestern University, Evanston, Illinois 60208

[30]University of Notre Dame, Notre Dame, Indiana 46556

[31]University of Oklahoma, Norman, Oklahoma 73019

[32]University of Panjab, Chandigarh 16-00-14, India

[33]Institute for High Energy Physics, 142-284 Protvino, Russia

[34]Purdue University, West Lafayette, Indiana 47907

[35]Rice University, Houston, Texas 77251

[36]University of Rochester, Rochester, New York 14627

[37]CEA, DAPNIA/Service de Physique des Particules, CE-SACLAY, France

[38]Seoul National University, Seoul, Korea

[39]State University of New York, Stony Brook, New York 11794

[40]SSC Laboratory, Dallas, Texas 75237

[41]Tata Institute of Fundamental Research, Colaba, Bombay 400005, India

[42]University of Texas, Arlington, Texas 76019

[43]Texas A&M University, College Station, Texas 77843

At $\sqrt{s} = 1.8$ TeV, production of $W$ and $Z$ bosons in $p\bar{p}$ collisions proceeds primarily via $q\bar{q}$ annihilation accompanied by an initial state gluon radiation which produces the transverse momentum, $p_T$, of $W$ and $Z$ bosons. Absolute predictions for the inclusive production cross sections, $\sigma_W$ and $\sigma_Z$, have been calculated to order $\alpha_s^2$ by van Neerven *et al.* [1]. In the low $p_T$ region ($p_T^W, p_T^Z < 20$ GeV/c) multiple soft gluon emission is expected to dominate the initial state radiation and the differential cross section, $d\sigma/dp_T$, has been calculated using a soft gluon resummation technique [2–6]. In the high $p_T$ region ($p_T^W, p_T^Z > 20$ GeV/c) perturbative QCD calculations are expected to describe the $d\sigma/dp_T$ [7]. Thus measurements of the inclusive and differential cross sections of $W$ and $Z$ production provide tests of QCD. Experimentally, use of leptonic decays of $W$ and $Z$ bosons, which do not involve final state strong interactions, allows for high precision measurements of their inclusive processes.

In this report we present measurements of the production cross sections times leptonic branching fractions and the $p_T$ distributions of $W$ and $Z$ bosons using data collected with the DØ detector [8] in the 1992–1993 run at the Fermilab Tevatron $p\bar{p}$ collider at $\sqrt{s} = 1.8$ TeV. We present preliminary results using a partial data sample from the 1994–1995 run. We also present a preliminary measurement of the $W$ charge asymmetry using $W \to \mu\nu$ sample, from which the information on parton distribution function (pdf) can be extracted.

## THE INCLUSIVE PRODUCTION CROSS SECTIONS [9]

The $W$ boson inclusive cross section is calculated as

$$\sigma_W \cdot B(W \to l\nu) = \frac{N_{obs} - N_{bkgd}}{\mathcal{A}_W \cdot \epsilon_W \cdot \mathcal{L}}, \tag{1}$$

where $N_{obs}$ is the number of observed events, $N_{bkgd}$ is the number of expected background events, $\mathcal{A}_W$ is the kinematic and geometric acceptance, $\epsilon_W$ is the detection efficiency, and $\mathcal{L}$ is the integrated luminosity used in the analysis. The $Z$ boson cross section, $\sigma_Z \cdot B(Z \to ll)$, is calculated in a similar fashion.

Electrons were detected in hermetic, uranium liquid-argon calorimeters with an energy resolution of about $15\%/\sqrt{E(\mathrm{GeV})}$. The central and end calorimeter regions were used in both the $W$ and $Z$ analyses, covering pseudorapidity ($\eta$) range: $|\eta| < 1.1$ and $1.5 < |\eta| < 2.5$, respectively. Muons were detected as tracks in three layers of proportional drift tube chambers outside the calorimeter: one 4-plane layer is located inside a magnetized iron toroid and two 3-plane layers are located outside, providing coverage for $|\eta| < 3.3$. The muon momentum resolution is $\sigma(1/p) = 0.18(p-2)/p^2 \ \oplus \ 0.008$ (with $p$ in GeV/c). Muons that passed through the central iron toroid ($|\eta| < 1.0$) were used in the cross section measurements. Neutrinos were inferred from the observed missing transverse energy ($\not{E}_T$) which was calculated using all the energy detected in the calorimeter cells out to pseudorapidity of 4.2. For the electron channel decays, the $\not{E}_T$ resolution was dominated by the underlying event and is $\sim 3$ GeV. For the muon channel decays, the muon transverse momentum was added to the calorimeter energy to calculate the total $\not{E}_T$, and the muon momentum resolution dominated the $\not{E}_T$ resolution.

TABLE I. Estimates of Backgrounds

| 1992–1993 data | $W \to e\nu$ | $Z \to ee$ | $W \to \mu\nu$ | $Z \to \mu\mu$ |
|---|---|---|---|---|
| $N_{\rm obs}$ | 10338 | 775 | 1665 | 77 |
| Backgrounds(%): | | | | |
| Multijet | $3.3 \pm 0.5$ | $2.8 \pm 1.4$ | $5.1 \pm 0.8$ | $2.6 \pm 0.8$ |
| $Z \to ee, \mu\mu, \tau\tau$ | $0.6 \pm 0.1$ | — | $7.3 \pm 0.5$ | $0.7 \pm 0.2$ |
| $W \to \tau\nu$ | $1.8 \pm 0.1$ | — | $5.9 \pm 0.5$ | — |
| Cosmic/Random | — | — | $3.8 \pm 1.6$ | $5.1 \pm 3.6$ |
| Drell-Yan | — | $1.2 \pm 0.1$ | — | $1.7 \pm 0.3$ |
| Total Background(%) | $5.7 \pm 0.5$ | $4.0 \pm 1.4$ | $22.1 \pm 1.9$ | $10.1 \pm 3.7$ |

The $W$ and $Z$ electron channel analyses based their event selection on a sample obtained with a single electron trigger ($Et > 20$ GeV). Offline, it was required that there be at least one electron with $E_T > 25$ GeV that passed "tight" electron identification cuts. Details of the electron identification are given in Ref. [10], with the main features being an electromagnetic (EM) cluster in the calorimeter with a matching track in the central tracking chambers. The electron was required to have the isolation variable less than 0.1, where the isolation is defined as $I$=($E_{\rm tot}(0.4)$-$E_{\rm EM}(0.2))/E_{\rm EM}(0.2)$, and $E_{\rm tot}(0.4)$ is the total calorimeter energy inside a cone of radius $\sqrt{\Delta\eta^2 + \Delta\phi^2} = 0.4$ and $E_{\rm EM}(0.2)$ is the electromagnetic energy inside a cone of 0.2. The cluster was also required to have transverse and longitudinal shapes consistent with those expected for an electron based on test beam measurements and Monte Carlo simulations. To select $W \to e\nu$ candidates, in addition to the "tight" electron with $E_T > 25$ GeV, events were required to have missing transverse energy $\not\!\!E_T > 25$ GeV. To select $Z \to ee$ candidates, in addition to the "tight" electron with $E_T > 25$ GeV, events were required to have a second electron with $E_T > 25$ GeV but the electron identification requirements were loosened by not requiring the track match in order to increase the efficiency. The invariant mass of the electron pair was required to be in the range $75 < M_{ee} < 105$ GeV/$c^2$. In an analysis of the 1992–1993 data sample, corresponding to $12.8 \pm 0.7$ pb$^{-1}$, we found 10338 $W$ and 775 $Z$ candidate events. The mass spectra for the $W \rightarrow e\nu$ and $Z \rightarrow ee$ events are shown in Fig. 1.

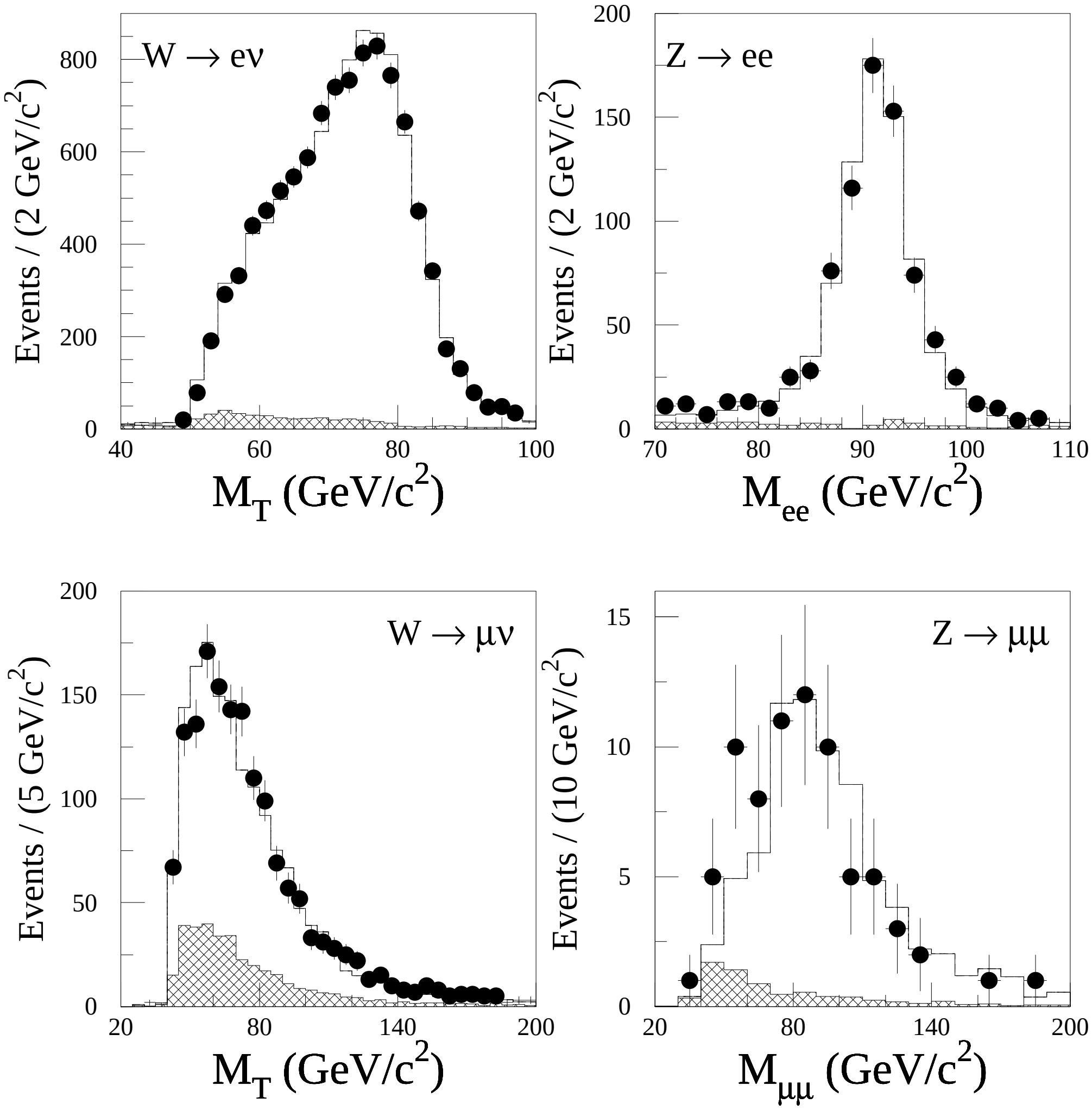

FIG. 1. Mass spectra from the 1992–1993 run. The points are the data, the shaded areas are the estimated backgrounds, and the histograms are the sum of the MC predictions and estimated backgrounds.

The muon channel $W$ and $Z$ analyses used an event sample obtained with a single muon trigger ($p_T > 15$ GeV). Offline, the events were required to have a reconstructed muon with $p_T > 20$ GeV. For $W \rightarrow \mu\nu$ events, the missing transverse energy was required to be $\not{E}_T > 20$ GeV. For $Z \rightarrow \mu\mu$ events, the offline threshold on the second muon was lowered to 15 GeV and the muon identification criteria were loosened. The main features of the muon identification (see Refs. [9,10] for details) include a good quality muon track that has a calorimeter confirmation signal and has a stringent match with a track in the central detector. Cosmic ray background was reduced by rejecting muons that also had hits or tracks within 10° in $\theta$ and 20° in $\phi$ in the muon chambers on the opposite side of the interaction point. For the $W \to \mu\nu$ selection, events that were $Z \to \mu\mu$ candidates were removed. From an analysis of the 1992–1993 data sample, corresponding to $11.4 \pm 0.6$ pb$^{-1}$, we found 1665 $W$ and 77 $Z$ candidate events. The observed mass spectra for the $W \to \mu\nu$ and $Z \to \mu\mu$ events are shown in Fig. 1.

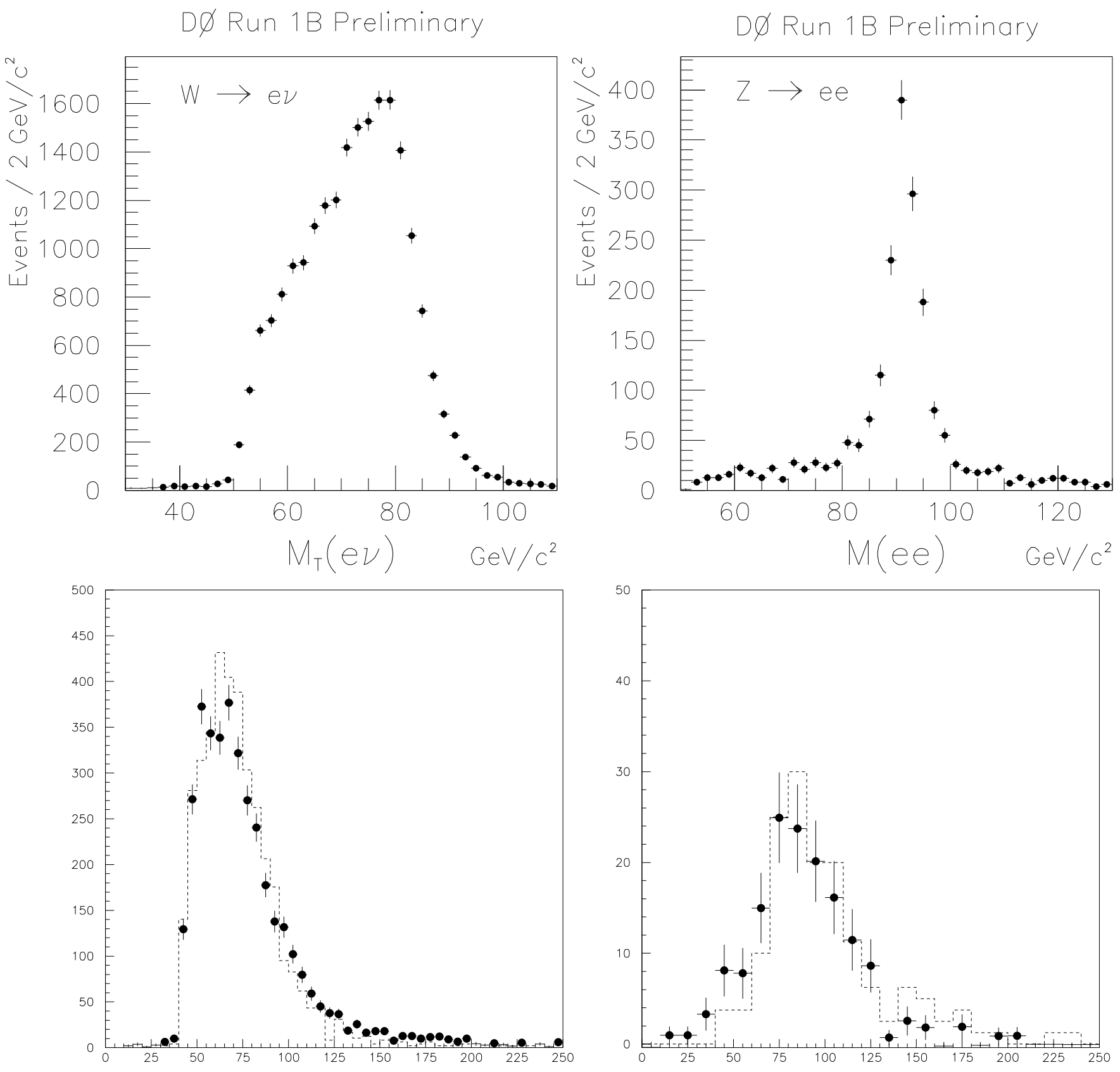

FIG. 2. Mass spectra (electron decay channels above, muon decay channels below) from a partial sample of the 1994–1995 data.

A preliminary analysis of a partial sample of the 1994–1995 data, corresponding to $25.1 \pm 1.4$ pb$^{-1}$, using the same requirements as described above, yielded 20998 $W \to e\nu$ and 1634 $Z \to ee$ candidates; an analysis of $30.7 \pm 1.7$pb$^{-1}$ yielded 4516 $W \to \mu\nu$ and 168 $Z \to \mu\mu$ candidates. The spectra are shown in Fig. 2.

TABLE II. Analysis results

| 1992–1993 | $W \to e\nu$ | $Z \to ee$ | $W \to \mu\nu$ | $Z \to \mu\mu$ |
|---|---|---|---|---|
| Nobs | 10388 | 775 | 1665 | 77 |
| Background (%) | $5.7 \pm 0.4$ | $4.0 \pm 1.4$ | $22.1 \pm 1.9$ | $10.1 \pm 3.7$ |
| Acceptance (%) | $46.0 \pm 0.6$ | $36.3 \pm 0.4$ | $24.8 \pm 0.7$ | $6.5 \pm 0.4$ |
| Efficiency (%) | $70.4 \pm 1.7$ | $73.6 \pm 2.4$ | $21.9 \pm 2.6$ | $52.7 \pm 4.9$ |
| $\mathcal{L}$ (pb$^{-1}$) | $12.8 \pm 0.7$ | $12.8 \pm 0.7$ | $11.4 \pm 0.6$ | $11.4 \pm 0.6$ |
| 1994–1995 *(Preliminary)* | $W \to e\nu$ | $Z \to ee$ | $W \to \mu\nu$ | $Z \to \mu\mu$ |
| Nobs | 20988 | 1634 | 4516 | 168 |
| Background (%) | $17.3 \pm 2.2$ | $11.0 \pm 2.4$ | $17.3 \pm 1.1$ | $10.1 \pm 3.7$ |
| Acceptance (%) | $46.1 \pm 0.6$ | $36.3 \pm 0.4$ | $22.0 \pm 0.9$ | $5.1 \pm 0.6$ |
| Efficiency (%) | $66.9 \pm 4.1$ | $70.6 \pm 4.6$ | $28.6 \pm 1.9$ | $60.9 \pm 2.6$ |
| $\mathcal{L}$ (pb$^{-1}$) | $25.1 \pm 1.4$ | $25.1 \pm 1.4$ | $30.7 \pm 1.7$ | $30.7 \pm 1.7$ |

The total backgrounds estimated for these event samples are shown in the spectra as hashed areas in Fig. 1 and are listed as a percentage of the observed number of events in Table I. A major background to the $W \to e\nu$ sample was from QCD multijet events where a jet was misidentified as an electron. It was estimated from data by measuring the $\not\!\!E_T$ distribution of a background-dominated sample, obtained by selecting events containing an EM cluster which failed at least one of the electron criteria (isolation, shower shape, and track match). We extrapolated this $\not\!\!E_T$ distribution into the signal region ($\not\!\!E_T > 25$ GeV) by normalizing the number of events in the background sample to that in the candidate sample (without the $\not\!\!E_T$ requirement imposed) in the region of small $\not\!\!E_T$ ($0 < \not\!\!E_T < 15$ GeV). The backgrounds due to $W \to \tau\nu \to e\nu\nu\nu$ decay and $Z \to e^+e^-$ where one of the electrons was lost, were estimated using a Monte Carlo simulation. The multijet background in the $Z \to ee$ sample was due to jet-jet or photon-jet events where the jets faked electrons in the detector. The amount of this background was estimated by fitting the invariant mass spectrum of the $Z \to ee$ events to the sum of the predicted $Z$ boson mass distribution and the experimentally determined multijet background shape. The invariant mass distributions for the jet-jet and photon-jet events were measured separately and then combined to obtain the overall multijet background shape.

The multijet background in the $W \to \mu\nu$ and $Z \to \mu^+\mu^-$ samples was estimated by comparing the distribution of energy in the calorimeter between the cones of radii of 0.2 and 0.6 around the muons with that measured for events containing a non-isolated muon and jets. The background in both the $W \to \mu\nu$ and $Z \to \mu^+\mu^-$ samples arising from $W \to \tau\nu$ and $Z \to \tau\tau$ decays, as well as the background in the $W \to \mu\nu$ sample arising from $Z \to \mu^+\mu^-$ where one of the muons was lost were estimated using a Monte Carlo simulation. The cosmic ray and random hit backgrounds to the $W \to \mu\nu$ and $Z \to \mu^+\mu^-$ samples were estimated from the distributions of muon time of origin relative to the beam crossing.

Finally, in determining the $Z \to ll$ cross section, a correction (which is listed as a background in Table I) was made for the Drell–Yan process where the lepton pair was produced via a virtual photon. This correction was sensitive to the choice of $Z$ mass window.

The kinematic and geometric acceptances (Tables II) were calculated using a Monte Carlo simulation which modeled the detector fiducial volume as well as the measured detector resolutions. The calculation used the CTEQ2M [11] pdf and a NLO calculation of $p_T^W$ and $p_T^Z$ by Arnold and Kauffman [5]. The largest contribution to the systematic error in the acceptance (Table II) arose from the choice of pdf. We estimated this uncertainty from the spread among the values obtained with CTEQ2 [11], MRS [12], and GRV [13]. Other errors included were from varying the $W$ mass, the simulation of the $p_T^W$ and $p_T^Z$ distributions, radiative corrections, the detector simulation of the $\not\!\!E_T$ distributions and the detector energy scale. The net detection efficiency (Table II) includes both the trigger and offline efficiencies. These were estimated from the data using $Z \to ll$ events since the trigger required only one lepton. The electron channel trigger was found to be $> 95\%$ efficient; and the muon trigger efficiency was 40% (70%) efficient for $W(Z)$ boson events.

The luminosity was measured by Level 0 trigger scintillator hodoscopes [14] mounted at $z = \pm 1.4$ m. The north-south coincidence rate was measured and corrected for multiple interactions. The visible cross section was calculated to be $\sigma_{L0} = 46.7 \pm 2.5$ mb, which resulted in a 5.4% relative error on the luminosity determination. This calculation was based on an average of the published CDF [15] and E710 [16] measurements of the total, elastic, and single diffractive cross sections, with the MBR [17] and Dual Parton Model DTUJET–93 [18] Monte Carlo routines used to determine the hodoscope acceptance.

TABLE III. Cross Section Results for electron ($e$) and muon ($\mu$) channels. When two errors are given the first is the statistical error and the second is total systematic error.

| | $\sigma_W \cdot B(W^\pm \rightarrow l^\pm \nu)$ (nb) | $\sigma_Z \cdot B(Z \rightarrow l^+ l^-)$ (nb) |
|---|---|---|
| 1992–1993 | | |
| $e$ | $2.36 \pm 0.02 \pm 0.15$ | $0.218 \pm 0.008 \pm 0.014$ |
| $\mu$ | $2.09 \pm 0.06 \pm 0.25$ | $0.178 \pm 0.022 \pm 0.023$ |
| 1994–1995 (Preliminary) | | |
| $e$ | $2.24 \pm 0.02 \pm 0.20$ | $0.226 \pm 0.006 \pm 0.021$ |
| $\mu$ | $1.93 \pm 0.04 \pm 0.20$ | $0.159 \pm 0.014 \pm 0.022$ |
| **Standard Model** | $2.42^{+0.13}_{-0.11}$ | $0.226^{+0.011}_{-0.009}$ |

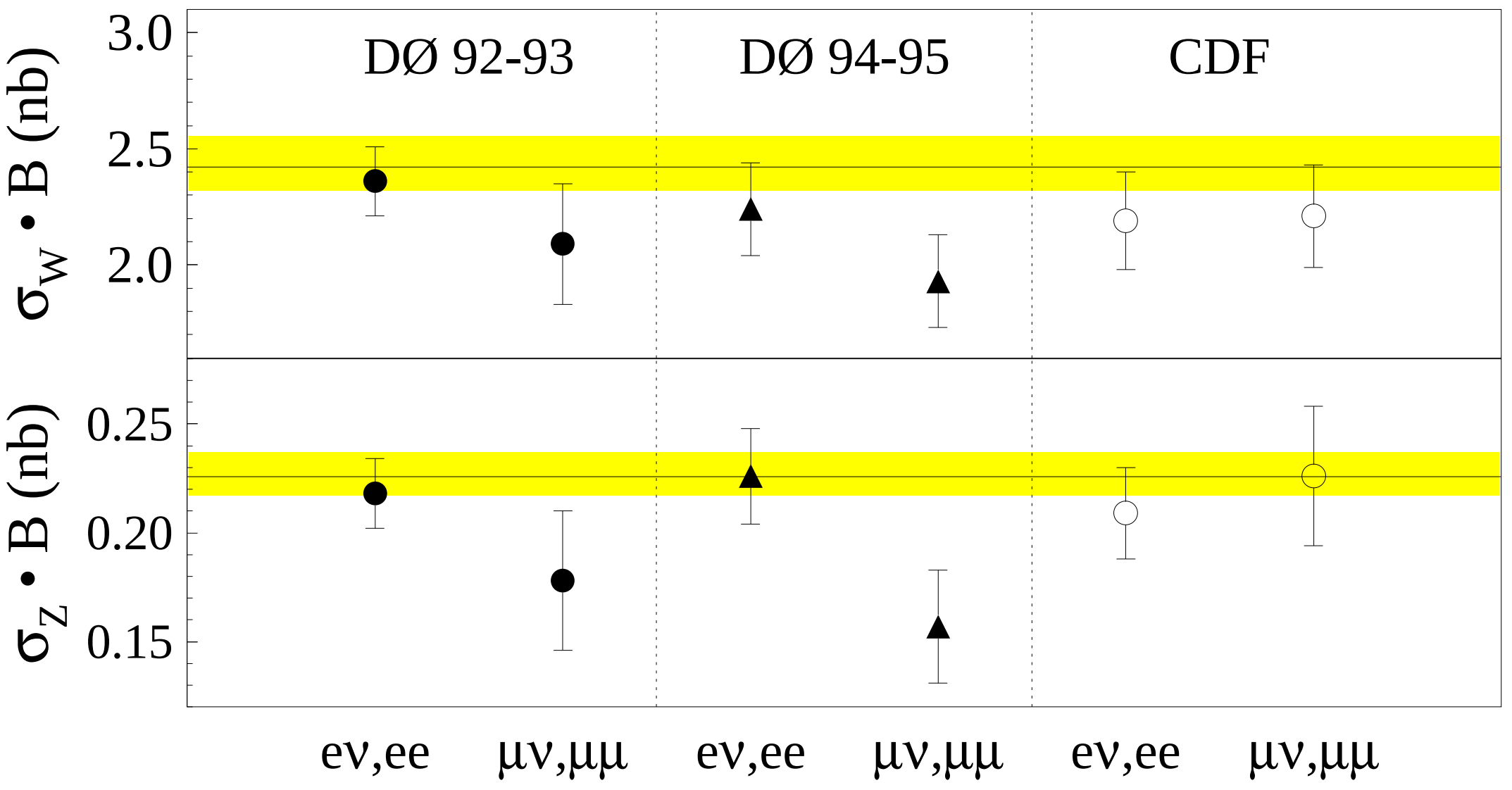

FIG. 3. $\sigma \cdot B$ for inclusive $W$ and $Z$ boson production. DØ 1994-1995 results are preliminary. The error bars indicate the combined statistical and systematic errors including the luminosity uncertainty. The solid lines are the predicted values calculated using the CTEQ2M pdf and the shaded bands indicate the uncertainty in the predictions.

The resulting cross sections, calculated using Eq. 1, are listed in Table III, where the first error given is statistical and the second is the total systematic error, including the luminosity uncertainty. These values are compared to the theoretical prediction (taken from Ref. [9]) in Fig. 3, together with the CDF results [19]. The total cross sections were calculated to be $\sigma_W = 22.35$ nb and $\sigma_Z = 6.708$ nb using a numerical calculation program from Ref. [1] and using the CTEQ2M pdf [11], $M_Z = 91.19$ GeV/c$^2$ [20], $M_W = 80.23 \pm 0.18$ GeV/c$^2$ [21], and $\sin^2\theta_W \equiv 1 - (M_W/M_Z)^2 = 0.2259$. The branching ratios used are $B(W \to l\nu) = (10.84 \pm 0.02)\%$ (calculated following Ref. [22] but with the above $M_W$), and $B(Z \to ll) = (3.367 \pm 0.006)\%$ [20]. The width of the band in Fig. 3 indicates the error in the predicted value, due primarily to the choice of pdf (4.5%) and to the use of a NLO pdf with the NLLO calculation (3%) [23]. Figure 4 shows $\sigma \cdot B$ for inclusive $W$ and $Z$ boson production cross sections from the DØ 1992-1993 data, the CDF measurements [19] and the measurements at the CERN $p\bar{p}$ collider [24] as a function of center of mass energy. Good agreement between the theoretical prediction and the measurements represents a success of perturbative QCD calculations.

## THE TRANSVERSE MOMENTUM DISTRIBUTION

The large $W/Z$ data samples collected during the 1992–1993 run allow significant improvement in the precision of the $d\sigma/dp_T$ measurement over previous measurements [25]. In this section we describe new, high precision measurements of the $p_T$ distributions of $W$ and $Z$ bosons. The data samples used for the $p_T^W$ and $p_T^Z$ measurements are identical to the sample described above for the $\sigma \cdot B(W \to e\nu)$ and $\sigma \cdot B(Z \to ee)$ measurements. The $p_T^W$ was determined from the hadronic recoil of the $W$, while the $p_T^Z$ was determined from the sum of two electron transverse momenta.

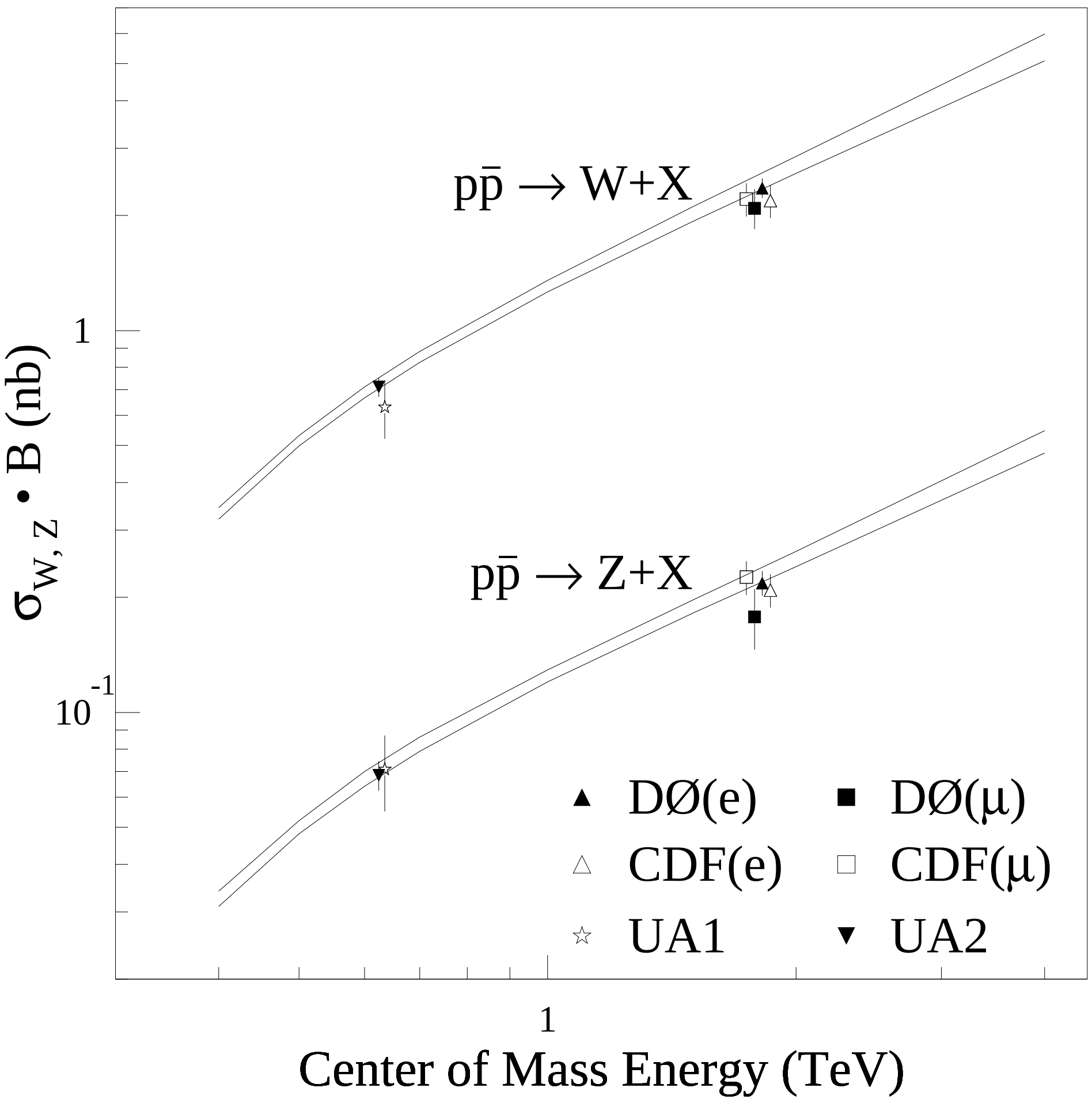

FIG. 4. $\sigma \cdot B$ for inclusive $W$ and $Z$ boson production as a function of center of mass energy. The error bars indicate the combined statistical and systematic errors including the luminosity uncertainty. The solid lines indicate the uncertainty of the Standard Model predictions.

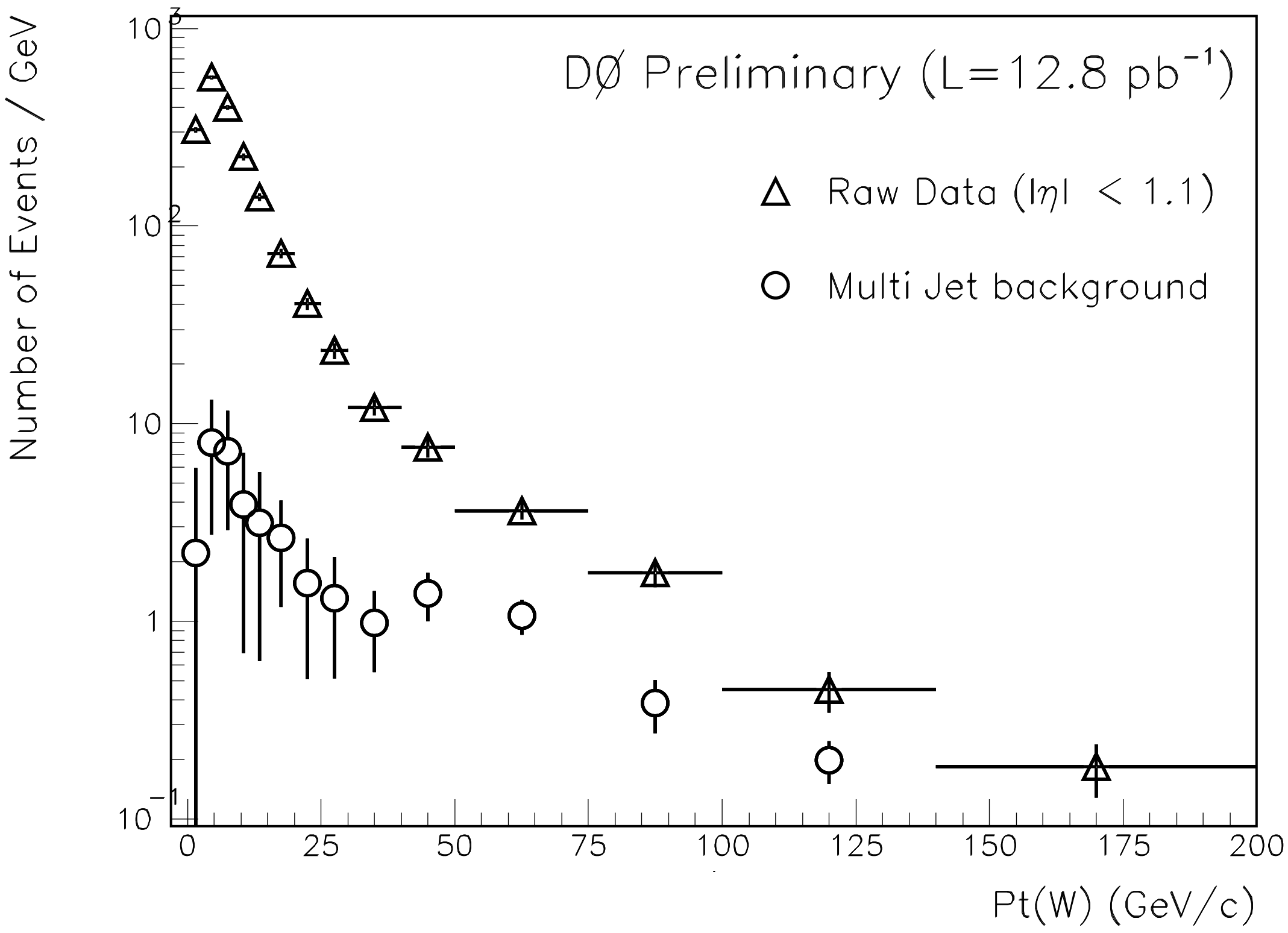

FIG. 5. Raw $p_T^W$ distribution (triangles) and multijet background (circles).

The measurement of the $p_T$ distribution requires the knowledge of the total amount of background, which is listed in Table I, and its shape as a function of the boson $p_T$. The shape of the background in the $W \to e\nu$ sample was obtained by subtracting the $p_T^W$ distribution obtained for a set of very clean electron identification cuts from a $p_T$ distribution of background-dominated sample while accounting for the relative efficiency loss between the two cuts. Figure 5 shows the obtained distribution superimposed on the raw $p_T^W$ distribution. The shape of the multijet background in the $Z \to ee$ sample was obtained from data by studying the product of the isolation variables of the two electrons as a function of the $e^+e^-$ invariant mass, shown in Fig. 6 for the standard and loose electron identification requirements. The events from region B ($75 < M_{ee} < 105$ GeV/c$^2$, and Iso1 $\cdot$ Iso2 $> 0.006$), marked in Fig. 6, were used to parametrize the shape of the multijet background.

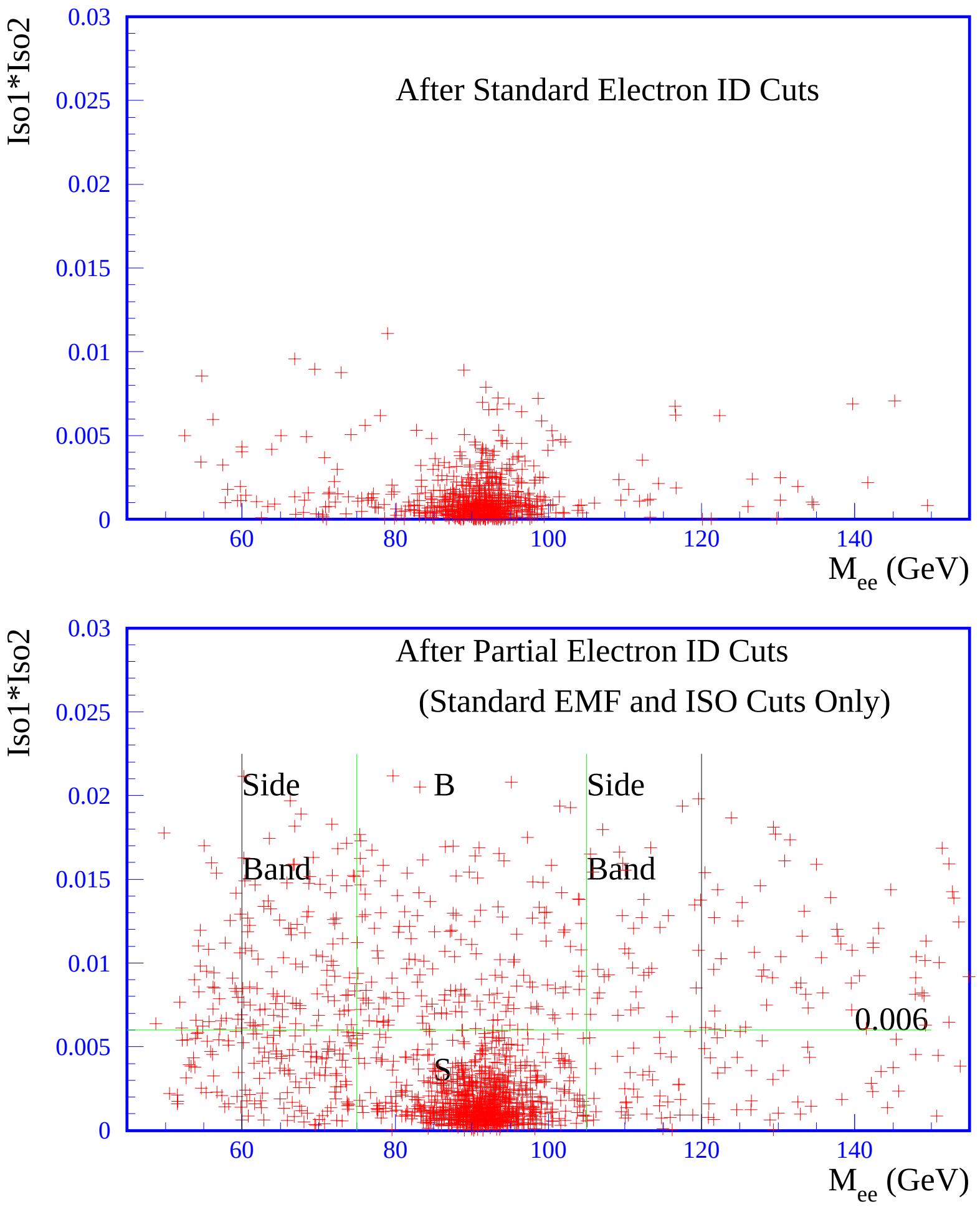

FIG. 6. Product of isolation variables of the two $Z$ boson electrons versus invariant mass, for the standard (top) and loose (bottom) electron identification requirements.

The systematic uncertainties of the $p_T^W$ measurement arose from uncertainties of: *i*) hadronic energy scale, *ii*) underlying event contribution, *iii*) hadronic resolution, and *iv*) background shape and magnitude. The hadronic energy scale was determined by balancing the $Z$ boson $p_T$ determined from the hadronic recoil and from the transverse momenta of the two electrons along the bisector of the angle subtended by the two. The uncertainty on the hadronic scale was thus controlled by the number of observed $Z$ candidates and for this data sample it produced $\sim 20\%$ uncertainty in the measurement. The magnitude of the underlying event contribution was also obtained from the $Z$ sample by matching the $Z$ boson $p_T$ resolution between data and Monte Carlo and was estimated to be of the order of 10% for this measurement. The uncertainty of the background shape and the magnitude was small in the low $p_T$ region but dominant in the high $p_T$ region. The statistical uncertainty for this sample was of the order of 5% per bin (high momentum bins had larger uncertainty, 10 – 30%). Thus the uncertainty of the $p_T^W$ measurement was dominated by systematic effects, most of which were directly controlled by the number of observed $Z$ bosons.

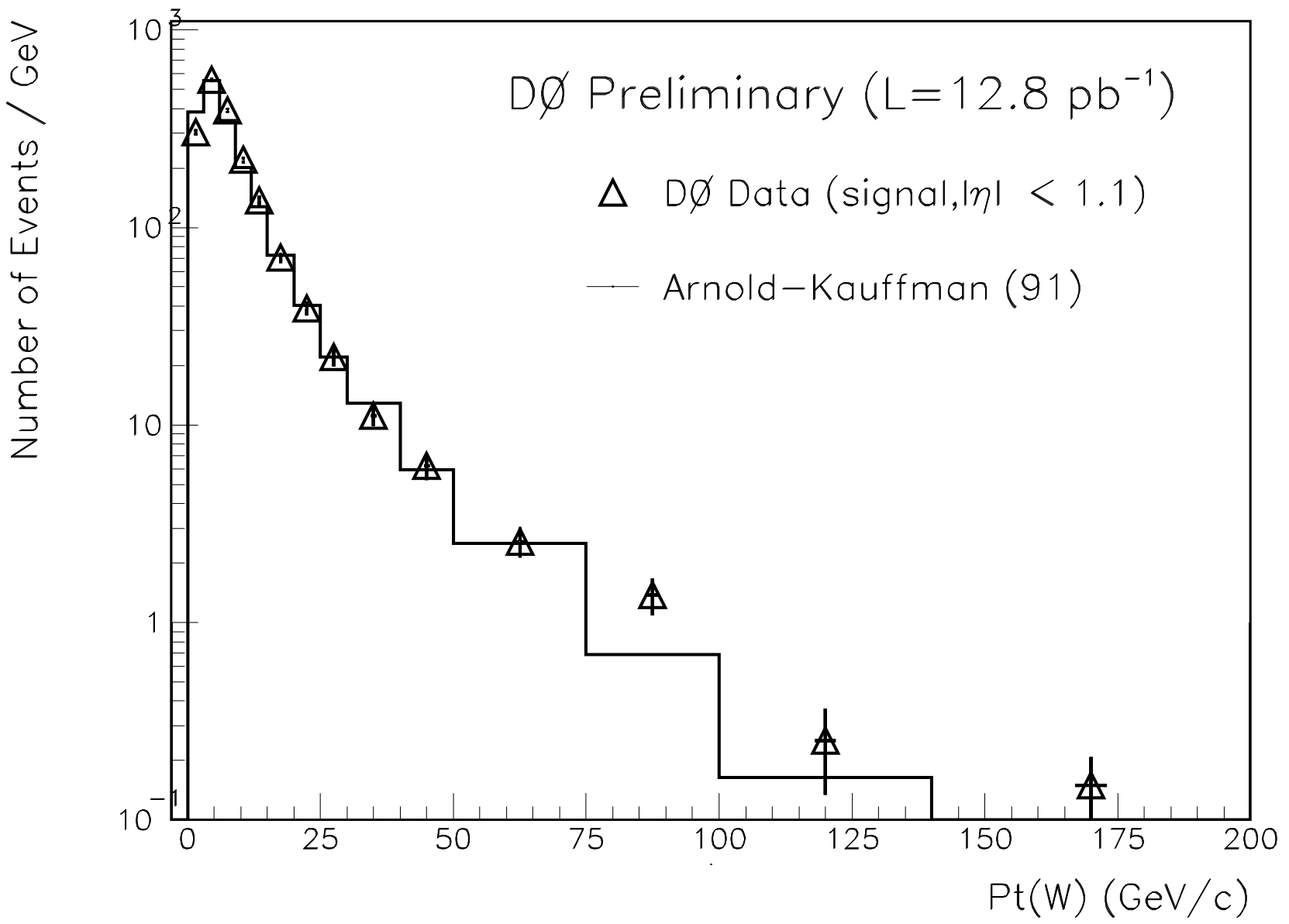

FIG. 7. Background subtracted $p_T^W$ distribution of data for $\mid \eta \mid < 1.1$ (triangles) with smeared theoretical prediction [5] (histogram).

The statistical uncertainty of the $p_T^Z$ measurement was of the order of $(10 - 20)\%$ up to 70 GeV/c. The systematic uncertainties arose from uncertainties of: *i*) electron energy scale, *ii*) electron energy resolution, and *iii*) electron angular ($\theta$ and $\phi$) resolutions. In the present analysis, the statistical uncertainty is the dominant uncertainty of the $p_T^Z$ measurement.

Figures 7 and 8 show the comparison between the background subtracted $p_T^W$ and $p_T^Z$ distributions, respectively, with theoretical predictions smeared by detector resolutions. The data tend to peak at a slightly higher value of $p_T$ than do the predictions.

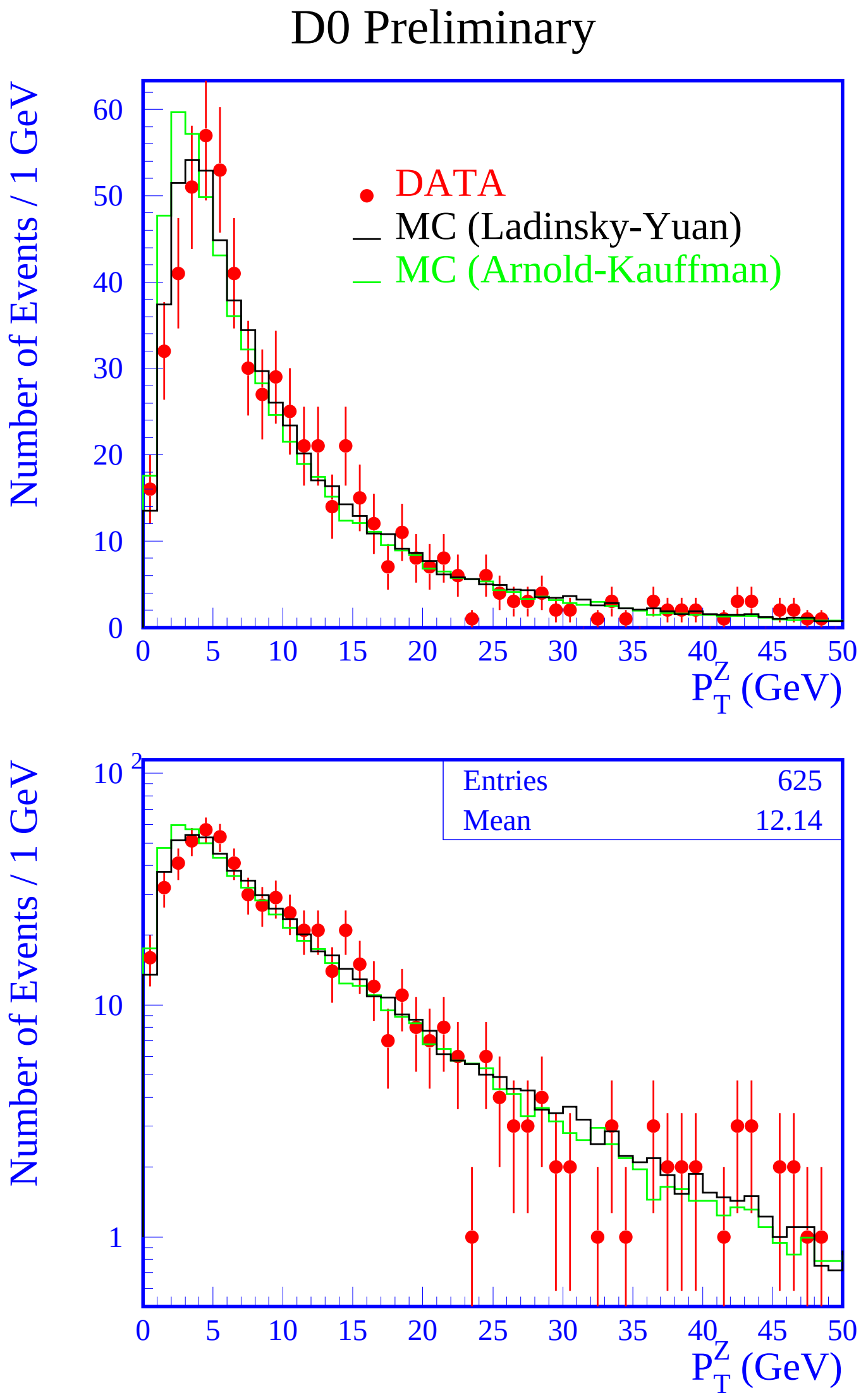

FIG. 8. Background subtracted $p_T^Z$ distribution (solid dots) with smeared theoretical predictions by [5] (gray histogram) and [6] (dark histogram) superimposed. Top plot: linear scale; Bottom plot: logarithmic scale.

## THE $W$ CHARGE ASYMMETRY

The $W$ production in $\bar{p}p$ collisions at $\sqrt{s} = 1.8$ TeV is dominantly from a valence-valence or valence-sea quark-antiquark interaction. Therefore a $W^+(W^-)$ is produced primarily by the interaction of a $u(d)$ quark from the proton and a $\bar{d}(\bar{u})$ quark from the antiproton. In the proton the $u$ valence quark momentum distribution, $u(x)$, is harder than the $d$ valence quark distribution, $d(x)$ and, therefore, a $W^+(W^-)$ is produced with a boost in the proton (antiproton) direction. Thus a measurement of the $W^+$ and $W^-$ rapidity distributions ($Y_{W+-}$) gives information on parton distribution function (pdf) in the region of low $x$ and high $q^2(\sim M_W^2)$ [26,27]. Because there is a twofold ambiguity in reconstructing $Y_W$ in a $W \to \ell\nu$ decay (due to the fact that the component of neutrino momentum along the beam direction is not measured) we measure the $Y_W$ distribution indirectly via the charged lepton rapidity distribution ($Y_\ell$), which is a sum of he $W$ rapidity and the lepton rapidity ($Y_\ell^{\rm CM}$) in the $W$ rest frame: $Y_{\ell^+} = Y_{W^+} + Y_{\ell^+}^{\rm CM}$, where $Y_\ell^{\rm CM}$ is determined by the $V-A$ couplings. At $\sqrt{s} = 1.8$ TeV the asymmetry due to $u(x)$ and $d(x)$ is larger than that from the $V-A$ effect and of the opposite sign. The experimentally convenient quantity is the charge asymmetry of the lepton pseudorapidity distribution,

$$A(\eta) \equiv \frac{d\sigma(\ell^+)/d\eta - d\sigma(\ell^-)/d\eta}{d\sigma(\ell^+)/d\eta + d\sigma(\ell^-)/d\eta}, \tag{2}$$

because it is insensitive to acceptance corrections. Furthermore, because $A(-\eta) = -A(\eta)$ by CP invariance the result can be shown as $A(|\eta|)$. A measurement of $A(\eta)$ with $|\eta| <\sim 1.7$ can provide information about the pdfs in the region of $x \sim 0.007 - 0.24$.

We present a preliminary result of the $W$ charge asymmetry using $W \to \mu\nu$ decays observed from the 6.5 pb$^{-1}$ data of the 1992–1993 run and the first $\sim$ 30 pb$^{-1}$ data of the 1994–1995 run. The data sample was obtained with a single muon trigger: a muon with $|\eta| < 1.7$ and $p_T^\mu > 15$ GeV/c. Additional track quality cuts, identical to those in the inclusive $W$ cross section analysis, and $p_T^\mu > 20$ GeV/c were imposed offline. For the 1992–1993 run, 60% of the data were taken with the muon toroid polarity in the forward direction and the remaining with reversed polarity, while the polarity was flipped every week for the 1994–1995 run to minimize possible detector charge asymmetry effects.

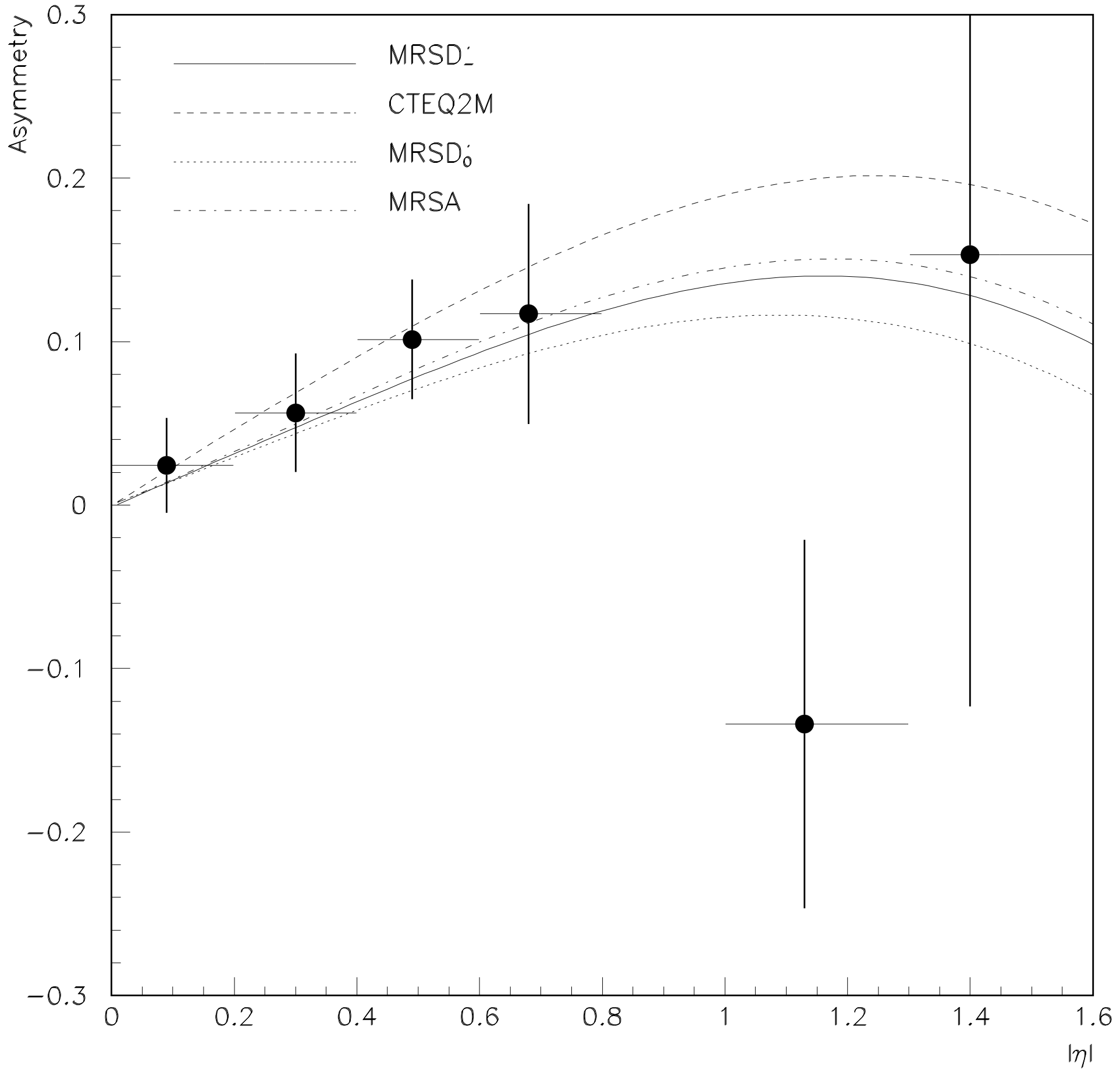

FIG. 9. DØ Preliminary $W$ decay muon charge asymmetry. The lines correspond to the theoretical predictions using several recent pdfs.

Muon charge misidentification dilutes the charge asymmetry. This systematic effect was estimated from the number of same sign pairs in the $Z \to \mu\mu$ sample. The charge misidentification probability was $8.8 \pm 5.2\%$ for the 1992–1993 run and $2.7 \pm 1.5\%$ for the 1994–1995 run. In addition, if the detector has different acceptance for $\mu^+$ and $\mu^-$, it can bias the charge asymmetry. Flipping the polarity of the muon toroid compensates this effect. The remaining uncompensated luminosity ($\sim 20\%$ in the 1992–1993 run and $\sim 1.5\%$ in the 1994–1995) was corrected for this bias using the factor derived from the data taken with magnet polarities in the forward and reverse directions.

Figure 9 shows a preliminary $W$ charge asymmetry measurement. Data are compared with a leading order calculation with input $p_T^W(y)$ spectrum obtained from the next-to-leading order resummation calculation of Ref. [6]. The data are consistent with the theoretical predictions with several recent pdfs.

## CONCLUSION

We have presented the inclusive production cross sections times leptonic branching fractions in both electron and muon channels. Good agreement between the theoretical calculation and our measurements indicates a success of perturbative QCD. We have shown the preliminary measurements of transverse momentum distributions of $W$ and $Z$ bosons in the electron channel. The large $W/Z$ data samples we obtained improve the precision of the $d\sigma/dp_T$ measurements over previous measurements. A preliminary measurement of the $W$ charge asymmetry using $W \to \mu\nu$ decays has also been presented.

## ACKNOWLEDGMENTS

We thank the Fermilab Accelerator, Computing, and Research Divisions, and the support staffs at the collaborating institutions for their contributions to the success of this work.

We also acknowledge the support of the U.S. Department of Energy, the U.S. National Science Foundation, the Commissariat à L'Energie Atomique in France, the Ministry for Atomic Energy and the Ministry of Science and Technology Policy in Russia, CNPq in Brazil, the Departments of Atomic Energy and Science and Education in India, Colciencias in Colombia, CONACyT in Mexico, the Ministry of Education, Research Foundation and KOSEF in Korea and the A.P. Sloan Foundation.

## REFERENCES

* Visitor from IHEP, Beijing, China.

‡ Visitor from CONICET, Argentina.

§ Visitor from Universidad de Buenos Aires, Argentina.

¶ Visitor from Univ. San Francisco de Quito, Ecuador.